# Eavesdropping Risk in Terahertz Channels by Covered Wavy Surfaces

Peian Li,[1,2] Wenbo Liu,[1,2] Jiabiao Zhao,[1,2] Jiayuan Cui,[1,2] Yapeng Ge,[1,2] Qiang Niu,[3] Yuping Yang,[3] Xiangzhu Meng,[4] and Jianjun Ma[1,2,5*]

[1]School of Integrated Circuits and Electronics, Beijing Institute of Technology, Beijing 100081, China
[2]Beijing Key Laboratory of Millimeter and Terahertz Wave Technology, Beijing 100081, China
[3]College of Science, Minzu University of China, Beijing 100081, China
[4]School of Automation, Beijing Institute of Technology, Beijing 100081 China
[5]Tangshan Research Institute, BIT, Tangshan, Hebei 063099, China

**Abstract:** Terahertz communications offer unprecedented data rates for next-generation wireless networks but suffer blockage susceptibility that restrict coverage and introduce physical-layer security vulnerabilities. Non-line-of-sight relay schemes using metallic wavy surfaces (MWS) address coverage limitations but require concealment beneath indoor materials for practical deployment. This work investigates THz channel characteristics and security vulnerabilities when MWS surfaces are covered with wallpaper, curtain, and wall plaster across 113-170 GHz. Results reveal that covering materials redistribute rather than eliminate eavesdropping threats, with persistent feasible interception scenarios remaining undetectable through conventional backscattering monitoring. These findings underscore the need for enhanced mechanisms designed for covered reflecting elements.

1. Introduction

Terahertz (THz) communication has emerged as a potential technology for next-generation wireless networks, offering unprecedented ultra-high data rates and exceptionally low latency capabilities that are essential for future applications [1], such as holographic communications, ultra-high-definition multimedia streaming, and real-time industrial automation systems. However, their deployment faces significant technical challenges that stem from the fundamental propagation characteristics - inherently high directionality and weak diffraction capability, which make them extremely susceptible to blockage and scattering effects caused by various environmental obstacles and surface irregularities [2, 3]. These propagation limitations severely restrict channel coverage and simultaneously introduce complex vulnerabilities to physical-layer security, creating substantial barriers to the practical implementation of THz wireless networks [4, 5]. To address these coverage limitations, the research community has developed several innovative non-line-of-sight (NLoS) relay schemes, such as reconfigurable intelligent surfaces (RIS) enabling real-time beam reconfiguration [6, 7], passive surfaces engineered with specific scattering characteristics [8-10], and unmanned aerial vehicles (UAVs) [11, 12] carry these RIS modules or reflector systems [13, 14].

While these NLoS relaying techniques substantially improve THz channel coverage and network accessibility, their implementation simultaneously introduces new categories of vulnerabilities at the physical layer. RIS-assisted channels exhibit heightened sensitivity to beam misalignment errors and phase control inaccuracies, where even minor deviations can result in unintended signal leakage toward unauthorized directions, thereby creating opportunities for malicious eavesdropping by adversarial users [15, 16]. UAV-based relay systems, despite their valuable spatial flexibility for dynamic coverage optimization, are subject to positional uncertainties and pointing errors that become particularly pronounced during high-speed movement scenarios or when mechanical vibrations affect platform stability, potentially causing unintentional exposure of confidential communications to unauthorized receivers [12, 17]. Passive scattering surfaces, while requiring no active control mechanisms, present security challenges through their inherent scattering behavior patterns, which are governed by structural geometry and surface roughness characteristics that may inadvertently redirect THz channels toward adjacent regions, consequently increasing the probability of signal interception by eavesdroppers [4].

Nevertheless, certain structural configurations, such as specifically designed wavy architectures, may offer potential pathways to reduce successful interception probabilities while maintaining acceptable coverage performance [5]. An important consideration for practical deployment scenarios involves the reality that these engineered surfaces may be intentionally concealed beneath building surfaces or integrated within decorative architectural layers for aesthetic considerations or other functional requirements. This deployment approach significantly complicates the propagation characteristics of THz channels, particularly because the dimensions of typical covering materials often fall within the range of THz wavelengths [18, 19], leading to complex electromagnetic interactions that can substantially alter both coverage patterns and security properties in ways that remain poorly understood [20]. Consequently, comprehensive investigation of the scattering properties exhibited by such covered surfaces and the resulting implications for eavesdropping risks represents a critical research priority for ensuring secure and reliable THz network deployment in real-world environments.

Previous research efforts have established important foundational knowledge regarding THz channel propagation through various materials and surface configurations. Koch *et al.* conducted extensive investigations into the scattering performance and dielectric properties of 50 representative building surface materials [2], encompassing both single-layered and multi-layered surface configurations through systematic measurements using THz time-domain spectroscopy (THz-TDS) techniques. Rasilainen [21] and Ma [3, 22] performed detailed experimental studies focusing on the reflection and transmission characteristics by various glass and plastic materials, revealing obvious multipath scattering phenomena and documenting substantial reflection and penetration losses. However, the effects of covering materials placed over engineered scattering surfaces remain largely unexplored, despite their critical importance for practical deployment scenarios. These covering layers introduce additional complexity that can further degrade channel propagation while simultaneously creating new and potentially more severe risks at physical layer.

To address this research gap and examine these effects under realistic deployment conditions, this work presents an investigation using representative indoor covering materials including wallpaper, curtain and wall plaster applied over metallic wavy surfaces (MWS). The selected MWS configuration has been previously demonstrated for passive channel beam allocation applications, with preliminary eavesdropping risk assessments establishing baseline security characteristics [5]. Through systematic experimental analysis, this work aims to quantify the impact of common indoor covering materials on both channel propagation performance and physical-layer security vulnerabilities, providing essential insights for the practical implementation of secure THz communication systems in realistic indoor environments.

2. **Experiment Setup**

The experimental validation employs a NLoS transmission testbed as illustrated in Fig. 1(a). The system integrates a THz transceiver configuration, metallic wavy surfaces covered with materials under test (MUT), and precision rotation stages, all assembled on a stabilized optical platform to minimize mechanical vibrations and ensure measurement accuracy. The transmitting subsystem utilizes a signal source (Ceyear 1465D) operating in continuous-wave (CW) mode with an initial output power of 0 dBm. The baseband signal undergoes frequency upconversion to the 113-170 GHz range through a high-performance frequency multiplier (Ceyear 82406B), enabling precise frequency control across this frequency band. The upconverted THz signal propagates through a standard horn antenna and is subsequently focused by a dielectric lens with a 10 cm focal length, creating a collimated beam with approximately 45 mm diameter at the target surface. This beam sizing ensures complete illumination of the test surface while maintaining sufficient power density for accurate measurements. The receiving subsystem employs an identical horn antenna configuration to maintain consistent beam characteristics and minimize measurement uncertainties. The received THz signal is captured and quantified using a calibrated power sensor (Ceyear 71718), providing reliable power measurements across the experimental frequency range. The experimental campaign encompasses three discrete frequencies - 113 GHz, 140 GHz, and 170 GHz - selected to characterize frequency-dependent scattering behavior.

To establish controlled NLoS propagation channels, precision-engineered metallic wavy surfaces (MWSs) are strategically positioned between the transmitter (Tx) and receiver (Rx) over a total path length of 68 cm. The MWS design leverages validated scattering characteristics [5], providing predictable and repeatable THz channel redirection under obstructed line-of-sight conditions. The MWS features a one-dimensional cosinoidal corrugation profile engineered for optimal angular scattering performance. Based on preliminary parametric studies [5], this investigation employs surfaces with a ripple amplitude of 0.7 mm and spatial period of 6 mm, parameters that maximize angular scattering efficiency while maintaining manufacturable tolerances. The MWS was precision-machined from high-purity aluminum plates with dimensions of 100 mm × 100 mm × 5 mm, providing adequate surface area to fully encompass the THz beam footprint and sufficient thickness to prevent unwanted transmission effects. Critical surface finishing includes precision polishing to achieve an average surface roughness of ~ 20 μm, a parameter essential for maintaining specular reflection characteristics in the THz frequency range. The MWS was positioned with careful consideration of surrounding metallic structures to minimize parasitic reflections and ensure measurement integrity. Operating as purely passive reflective elements, the MWS exploits its geometric corrugation combined with aluminum's high conductivity [23, 24] to create controlled scattering patterns, enabling systematic investigation of THz channel propagation behavior in obstructed environments.

Building upon the baseline MWS characteristics, this work extends to examine the influence of practical surface coverings by applying three representative indoor construction materials: wallpaper, curtain, and wall plaster. These materials represent common architectural elements that may overlay engineered surfaces in real-world deployments. The wallpaper and curtain samples are applied using precision adhesive mounting techniques. However, due to the underlying cosinoidal profile of the MWSs, unavoidable air gaps form between the covering materials and the metallic substrate, creating complex multi-layer electromagnetic interfaces. In contrast, wall plaster - primarily composed of titanium dioxide with controlled particle size distribution [25] - is applied in wet paste form and uniformly spread across the surface using standard construction techniques. The plaster undergoes controlled air-drying followed by precision sanding to achieve a smooth, flat interface that effectively eliminates the underlying corrugated geometry, as demonstrated in Fig. 1(b). Three wallpaper samples (P1, P2, P3) with varying surface roughness characteristics are selected for comprehensive analysis. All them are fabricated from polyvinyl chloride (PVC)

[26] with thicknesses ranging from 0.22 to 0.25 mm, ensuring comparable material properties while providing different surface morphologies. Sample P1 exhibits quasi-regular, densely distributed surface ridges creating periodic texture variations. Sample P2 features non-uniform stripe patterns with irregular geometries that introduce random scattering elements. Sample P3 presents a smooth surface with negligible texture variations, serving as a baseline for comparison. The curtain sample consists of polyester fabric [27] with a tightly woven fiber structure exhibiting nearly smooth surface morphology and measured thickness of 0.64 mm. The knitted mesh architecture creates a porous medium that influences THz channel interaction through both surface scattering and partial transmission effects.

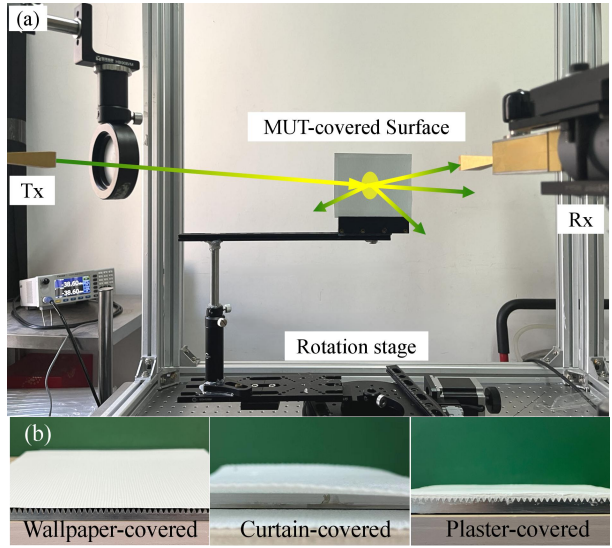

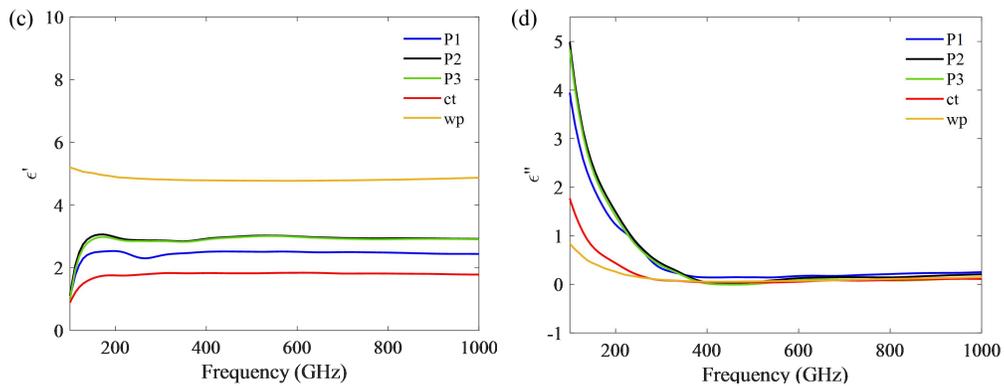

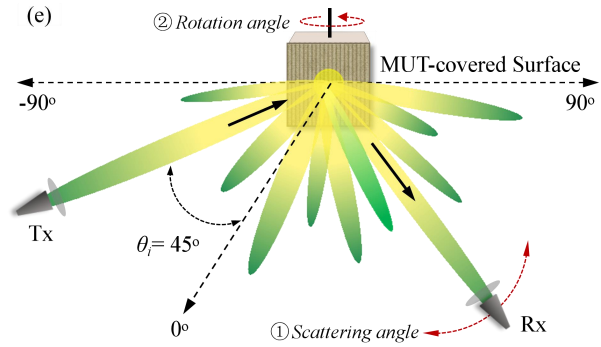

Fig. 1. (a) NLoS THz transmission testbed with material-covered metallic wavy surfaces and dual rotation stages. (b) Cross-sectional views of wallpaper, curtain, and wall plaster surface coverings. (c,d) Real and imaginary permittivity of covering materials measured by THz-TDS. (e) Dual-mode measurement protocol for angular scattering and incident angle characterization.

The material characterization employs the T-SPEC 800 THz-TDS system operating across a 0.1-3 THz bandwidth. This system incorporates a voice coil-driven fast delay line enabling temporal waveform acquisition up to 116 ps within 0.1 seconds, providing spectral resolution of approximately 8 GHz [28]. The high temporal resolution facilitates accurate extraction of material parameters despite sample thickness variations. Fig. 1(c) presents the real part of permittivity ($\varepsilon'$) across the measurement frequency range. The three wallpaper samples demonstrate comparable $\varepsilon'$ values with minor variations attributable to manufacturing tolerances and surface texture differences. The curtain sample exhibits much lower $\varepsilon'$ values, corresponding to reduced reflectivity and enhanced transmission characteristics. These differences arise from both material composition and structural density variations, where the curtain's knitted mesh structure facilitates THz wave penetration compared to the solid PVC wallpaper construction. The imaginary part of permittivity ($\varepsilon''$) evolution is illustrated in Fig. 1(d), revealing distinct frequency-dependent absorption characteristics. The wallpaper samples exhibit elevated $\varepsilon''$ values at lower frequencies that decrease rapidly with increasing frequency, stabilizing above 400 GHz. This behavior indicates strong frequency-dependent absorption mechanisms typical of polymer materials in the THz range. The curtain sample demonstrates slightly lower $\varepsilon''$ values across the frequency range, indicating weaker absorption properties. The wall plaster sample exhibits the lowest $\varepsilon''$ values throughout the measurement bandwidth, suggesting minimal dielectric losses and predominantly reflective behavior. For all the samples, the non-uniform surface morphology introduces local thickness variations that can affect permittivity calculations. To mitigate these effects, measurements are conducted at multiple locations across each sample, and averaged values are used for analysis, ensuring representative material parameter extraction.

The experimental protocol incorporates dual rotational measurement modes to characterize directional scattering behavior and signal leakage pathways under NLoS conditions, as illustrated in Fig. 1(e). This methodology enables systematic evaluation of both angular power distribution and incident angle dependencies. In the first measurement mode (Mode ①), the receiver is mounted on a precision motorized rotation stage while maintaining the transmitter and test surface in fixed positions. The receiver traverses a circular trajectory centered on the test surface, preserving constant propagation path length throughout the angular sweep. This configuration facilitates measurement of scattered power distribution as a function of observation angle, providing critical data for eavesdropping vulnerability assessment. The second measurement mode (Mode ②) maintains both transmitter and receiver in stationary positions while rotating the test surface to vary the incident angle. This approach enables analysis of how scattered THz channels evolve with respect to surface orientation, revealing angle-dependent scattering characteristics crucial for understanding coverage patterns and security implications.

## 3. Channel performance and security analysis

### 3.1 Wallpaper-covered channel

Three wallpaper samples (P1, P2, P3) with distinct surface roughness characteristics were individually applied to the MWS following the configuration specified in Fig. 1. Angular distribution measurements employed the rotation protocol indicated by Mode ①, where the receiver traversed 90° in 1° increments while maintaining fixed transmitter and surface positions. The incident angle was set to 45°, establishing the specular reflection baseline for comparative analysis.

Figure 2(a) demonstrates the measured scattering patterns for both bare metallic surfaces and wallpaper-covered configurations at 140 GHz. A significant enhancement in channel power strength along the specular reflection direction is observed, attributed to constructive interference between two distinct propagation components: the specular reflection from the wallpaper surface (component ① in Fig. 2(c)) and the Bragg scattering component from the underlying metallic wavy substrate after transmission through the wallpaper layer (component ② in Fig. 2(c)). This dual-path constructive interference mechanism substantially improves channel performance in the specular direction by approximately 10 dB compared to bare surfaces. However, in non-specular directions, modest channel reductions occur due to penetration losses through the wallpaper dielectric (characterized by $\varepsilon''$ values in Fig. 1(d)) and additional scattering induced by surface texture variations. Critically, the angular positions of scattering lobes remain unchanged compared to the bare surface, confirming that the underlying periodic structure governs the scattering behavior according to Bragg's Law [5], as validated by theoretical predictions in the inset of Fig. 2(a). While wallpaper coverings provide effective concealment for engineered surfaces in practical deployments, this preservation of scattering directionality combined with enhanced specular performance paradoxically creates more favorable conditions for eavesdropping.

Quantitative security analysis employs the normalized *secrecy capacity metric* $c_{s\text{-}p}$ to evaluate eavesdropping vulnerability under wallpaper-covered conditions. The experimental scenario positions the transmitter (Alice) and legitimate receiver (Bob) at the optimal 45° incident/reflection angle pair, while the potential eavesdropper (Eve) is modeled as the rotating receiver capturing scattered signals across multiple angles. The *normalized secrecy capacity* [5] is defined as

$$c_{s-p} = \frac{\lg(1+SNR_{Bob}) - \lg(1+SNR_{Eve-p})}{\lg(1+SNR_{Bob})} \quad (1)$$

where, $SNR_{Bob}$ represents the signal-to-noise ratio at the legitimate receiver with uncovered surfaces, and $SNR_{Eve\text{-}p}$ denotes the SNR measured at each angular position with wallpaper coverings. A threshold of $c_{s\text{-}p}$<0.5 indicates feasible eavesdropping scenarios where Eve receives sufficiently strong signals for potential information extraction [4]. Fig. 2(b) reveals that despite wallpaper application, multiple angular directions maintain $c_{s\text{-}p}$<0.5 (highlighted in yellow regions), indicating persistent eavesdropping vulnerabilities. The wallpaper covering does not eliminate all security risks but narrows the threat profile while enhancing the specular channel.

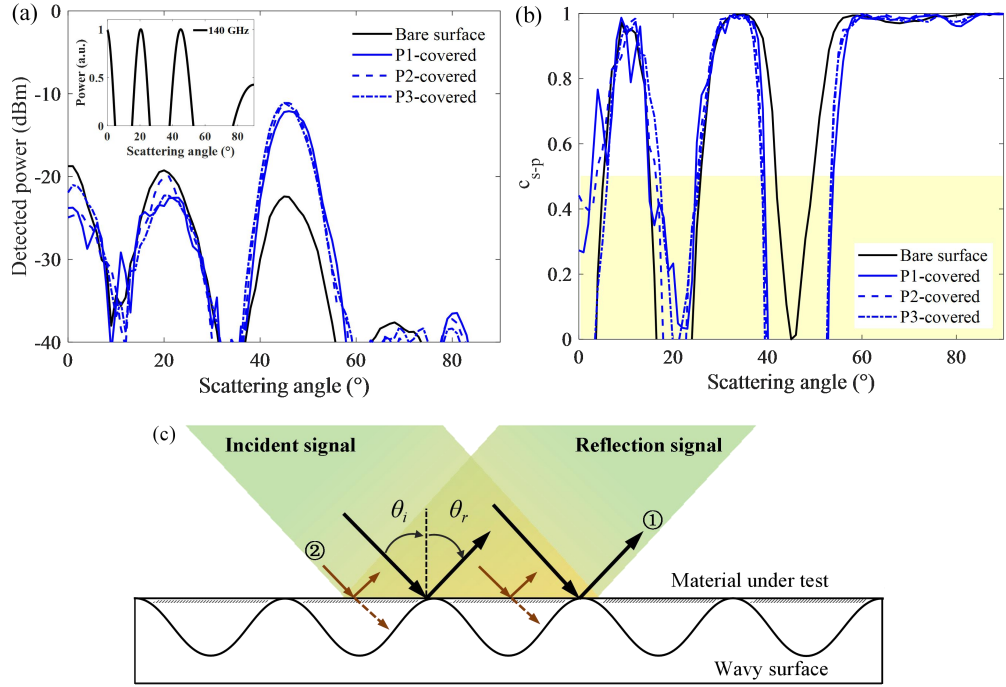

Fig. 2. Wallpaper-covered surface results at 140 GHz. (a) Angular scattered power distribution. Inset: theoretical predictions. (b) Normalized secrecy capacity distribution; yellow regions indicate $c_{s\text{-}p} < 0.5$ eavesdropping vulnerability. (c) Dual-path channel propagation schematic.

There is a critical consideration for practical eavesdropping scenarios involves surface manipulation to optimize signal interception while avoiding detection by the legitimate receiver. To evaluate this capability, Alice and Bob positions remain fixed while the surface undergoes controlled rotation following Mode ② in Fig. 1(e), with counterclockwise rotation defined as positive angular displacement. The *variation parameter* $v_{p\text{-}p}$ [5] quantifies signal fluctuations observable by Bob, as

$$v_{p-p} = 1 - \frac{SNR_{Bob-p}}{SNR_{Bob}} \qquad (2)$$

where, $SNR_{Bob\text{-}p}$ represents Bob's SNR with covered, rotated surfaces, and $SNR_{Bob}$ indicates the baseline SNR with bare surfaces at 45° incidence. The threshold of $v_{p\text{-}p}=0.5$ defines acceptable variation levels that remain undetectable to Bob, enabling successful eavesdropping operations.

Figure 3(a) illustrates the channel power variation as a function of surface rotation angle for both bare and wallpaper-covered configurations. At 0° rotation, Bob achieves maximum signal strength due to optimal specular alignment. As rotation increases, the incident angle deviation causes gradual departure from the specular path, reducing received power. The difference between covered and bare surfaces diminishes with increasing rotation due to attenuation of the wallpaper-specific specular component while penetration losses become proportionally more significant.

The analysis identifies specific rotation angles where $v_{p\text{-}p} < 0.5$, as shown in Fig. 3(b), indicating feasible eavesdropping manipulation scenarios where surface orientation can be optimized for signal interception while maintaining imperceptible variations from Bob's perspective. This frequency-dependent angular behavior is consistently observed across the experimental bandwidth, as demonstrated in Fig. S1(a) in the supplemental document, where scattering directions shift according to Bragg's Law for different operating frequencies (113 GHz and 170 GHz). Two representative surface rotation angles - 14° (59° incidence $\theta_i$) and 31° (76° incidence $\theta_i$) -were selected for detailed eavesdropping assessment based on their favorable concealment characteristics and comparable signal levels to Bob's reception. Crucially, these larger incidence angles reduce backscattering signatures toward the transmitter due to the angular dependence of radar cross-section, where off-normal incidence redirects reflected power away from the source direction [5], thereby enhancing covert interception capabilities. The optimal incidence angles vary with operating frequency due to the wavelength-dependent nature of the Bragg scattering condition, as indicated in Fig. S1(b) in the supplemental document. Fig. 3(c) presents the corresponding secrecy capacity distributions for both rotation configurations, revealing substantial angular ranges that satisfy $c_{s\text{-}p} < 0.5$ at both settings, confirming effective multi-directional eavesdropping opportunities. The distinct scattering characteristics under different incidence angles provide eavesdroppers with enhanced deployment flexibility and multiple interception vectors, enabling adaptive positioning strategies that exploit the frequency-selective nature of the wavy surface response.

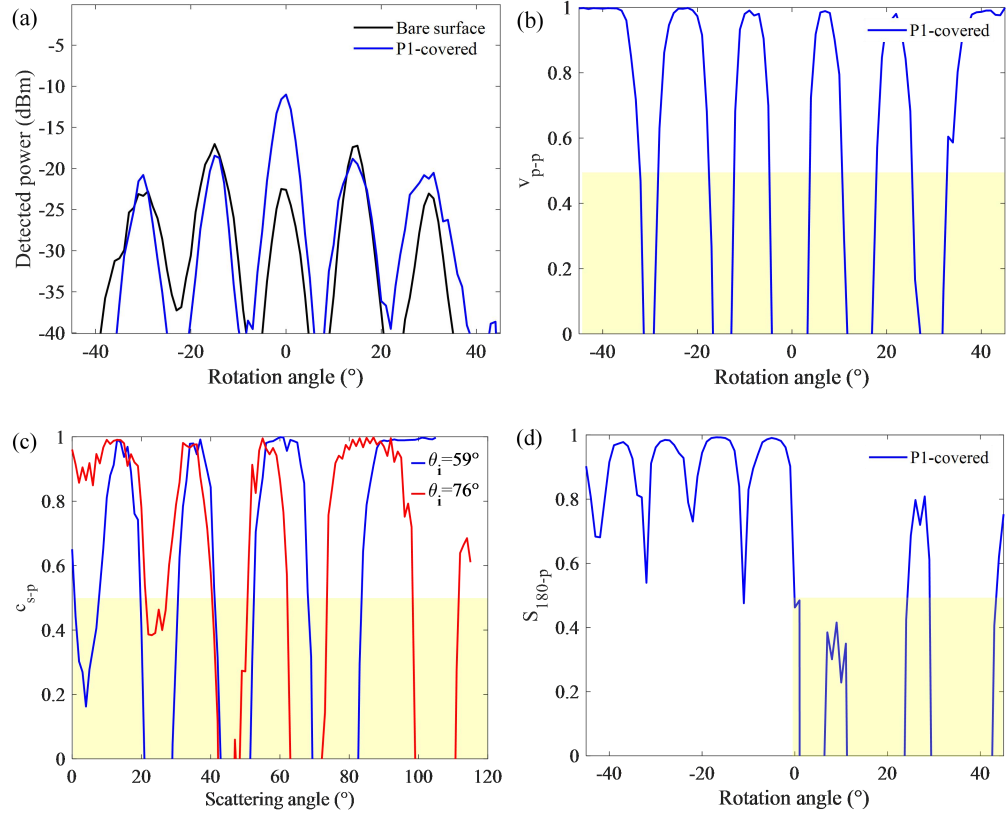

Fig. 3. Wallpaper security validation at 140 GHz. (a) channel power vs. surface rotation angle. (b) Variation parameter $v_{p\text{-}p}$ enabling covert manipulation where $v_{p\text{-}p} < 0.5$. (c) Secrecy capacity $c_{s\text{-}p}$ at 14° and 31° rotation angles. (d) Backscattering parameter $S_{180\text{-}p}$ distribution. Yellow regions: $c_{s\text{-}p} < 0.5$.

Backscattering analysis evaluates the transmitter (Alice)'s ability to detect anomalous reflectors through co-located transmitter-receiver measurements across the same angular range [29] as Fig. 3(a). The *backscattered variation* parameter $S_{180\text{-}p}$ [4] is defined as

$$S_{180-p} = \left| 1 - \frac{SNR_{Alice-p}^{nosurface}}{SNR_{Alice-p}^{surface}} \right| \quad (3)$$

where, $SNR_{Alice-p}^{nosurface}$ denotes the SNR at Alice in the absence of coverings, while $SNR_{Alice-p}^{surface}$ represents the SNR at Alice when the reflective surface is covered. Fig. 3(d) demonstrates that $S_{180\text{-}p}$ remains below the 0.5 detection threshold for rotation angles at 14° and 31° in Fig. 3(c), indicating insufficient backscattering variation to trigger security alerts. This resistance to backscattering detection mechanisms further enhances the covert operation capabilities of wallpaper-covered eavesdropping surfaces, as the wallpaper layer effectively masks the signature changes that might otherwise reveal the presence of manipulated reflecting elements.

*3.2 Curtain-covered channel*

Curtain material represents another common indoor covering with distinct electromagnetic properties compared to wallpaper. Our experimental validation employed the polyester curtain sample (0.64 mm thickness) applied to the MWS using the identical protocol established in Section 3.1. Fig. 4(a) demonstrates minimal impact on angular distribution patterns compared to uncovered surfaces. The preservation of scattering directionality stems from the curtain's low dielectric constant (ε′) and reduced reflectivity, as characterized in Fig. 1(c). Minor power variations result from weak reflection superposition effects and limited transmission losses through the curtain material. Power fluctuations near the specular direction are attributed to imperfect surface contact due to the curtain's flexible textile structure, which prevents complete adherence to the underlying corrugated geometry.

Eavesdropping vulnerability analysis employs the established *secrecy capacity metric* from Eq. (1). Fig. 4(b) reveals substantial angular regions satisfying $c_{s\text{-}p} < 0.5$, particularly near the specular direction and around scattering angles of 0° and 21°. These results indicate persistent interception feasibility despite curtain coverage, though the threat distribution differs from wallpaper configurations due to distinct material interactions. Specifically, the curtain's lower dielectric constant (ε′ ≈ 1.8 vs. 2.5 for wallpaper, Fig. 1(c)) and porous mesh structure create weaker surface reflections and enhanced transmission compared to the solid PVC wallpaper construction. This results in reduced constructive interference between surface and substrate reflections, leading to different angular vulnerability patterns where the curtain maintains more uniform threat distribution across multiple scattering directions rather than the concentrated enhancement observed with wallpaper coverings.

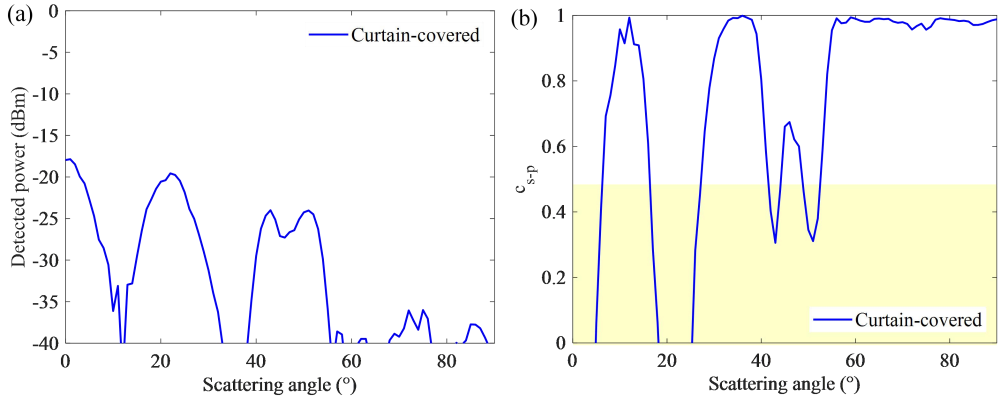

Fig. 4. Curtain-covered surface at 140 GHz. (a) Angular scattered power distribution with and without curtain coverage. Inset: theoretical model. (b) Secrecy capacity $c_{s\text{-}p}$ distribution with the curtain-covered surface; yellow regions indicate eavesdropping feasibility.

Figure 5(a) presents received power distributions across rotation angles, demonstrating minimal variation again due to the curtain's low reflectivity and negligible penetration losses. The *variation parameter* analysis in Fig. 5(b) identifies multiple rotation angles satisfying $v_{p\text{-}p}$ < 0.5, enabling covert surface manipulation without detection by the legitimate receiver. A specific rotation angle of 13° (58° incidence) was selected for detailed eavesdropping validation based on optimal concealment characteristics. Fig. 5(c) shows the corresponding $c_{s\text{-}p}$ distribution, revealing multiple non-specular directions suitable for signal interception. The backscattering performance in Fig. 5(d) confirms $S_{180\text{-}p}$ < 0.5 at the 13° rotation setting, indicating resistance to transmitter-based detection mechanisms.

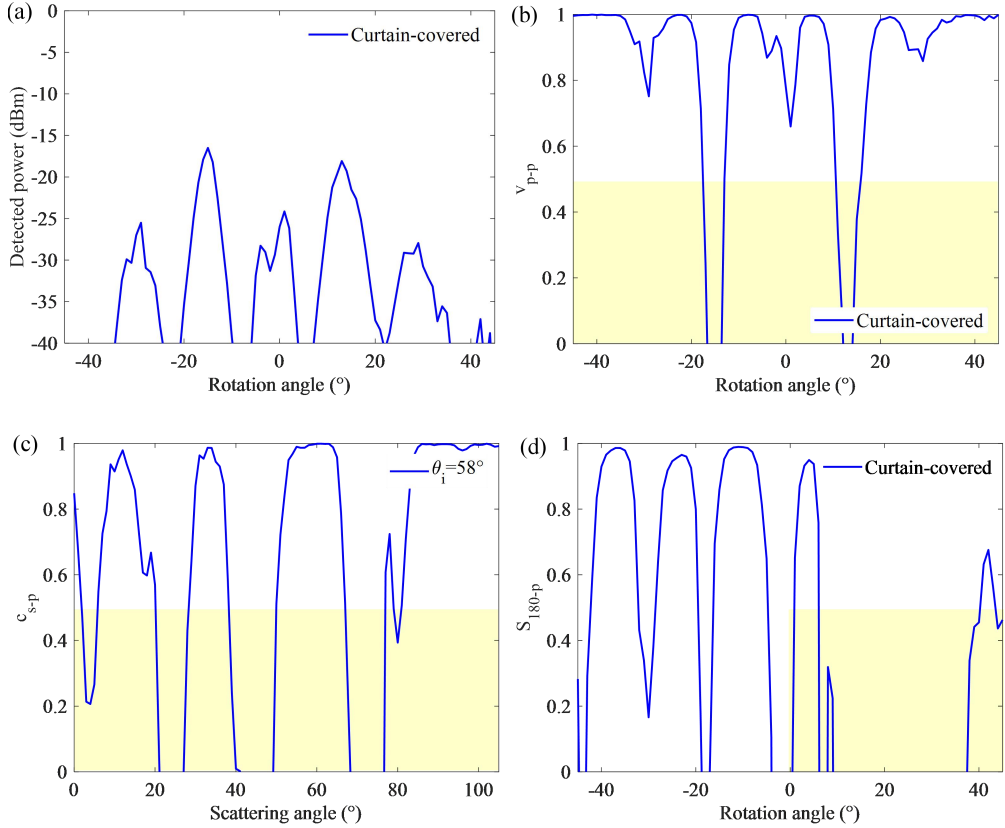

Fig. 5 Curtain security validation at 140 GHz. (a) Bob's signal vs. rotation angle. (b) Variation parameter $v_{p\text{-}p}$ distribution under different surface rotation angles. (c) Secrecy capacity $c_{s\text{-}p}$ at 13° rotation. (d) Backscattering parameter $S_{180\text{-}p}$ under different surface rotation angles with the curtain-covered surface. Yellow regions: $c_{s\text{-}p}$ < 0.5.

*3.3 Plaster-covered channel*

Unlike previous coverings (wallpaper and curtain), the titanium dioxide-based plaster fills the corrugated surface undulations, creating smooth interfaces with thickness variations between 0.6-2 mm due to the underlying surface profile. Fig. 6(a) presents scattering measurements at 140 GHz, showing preserved directional characteristics with enhanced specular reflection due to constructive interference. The relatively disordered angular distribution compared to flexible coverings results from the plaster's complete interfacial contact and filling of surface irregularities. Local thickness variations introduce boundary condition changes that contribute to additional phase disturbances and secondary scattering effects. The *secrecy capacity* performance in Fig. 6(b) reveals fewer angular positions satisfying $c_{s\text{-}p}$ < 0.5 compared to wallpaper or curtain configurations, attributed to the plaster's significantly higher dielectric constant (see Fig. 1(c)) and resulting stronger electromagnetic interactions that create more complex interference patterns [2], effectively dispersing scattered power across wider angular ranges and reducing concentrated vulnerability zones. However, sufficient interception opportunities remain available for potential eavesdroppers.

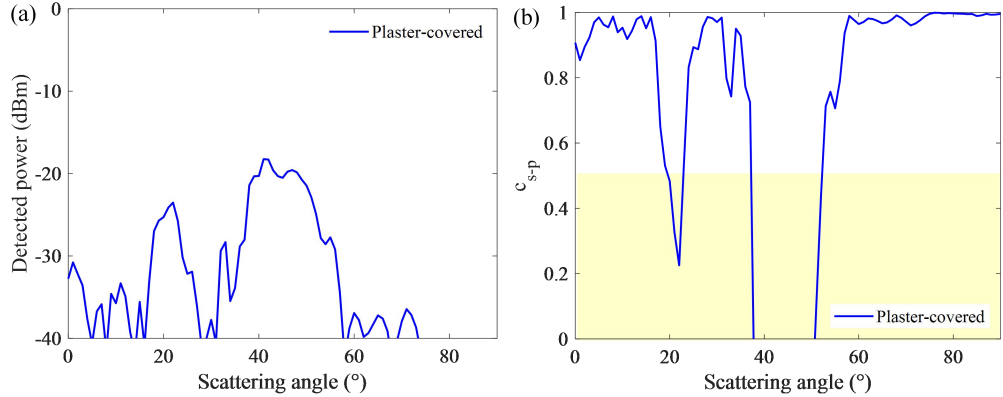

Fig. 6. Wall plaster results at 140 GHz. (a) Angular scattered power with enhanced specular reflection. (b) Secrecy capacity $c_{s-p}$ showing reduced but persistent vulnerabilities; yellow regions indicate $c_{s-p} < 0.5$.

Figure 7(a) demonstrates Bob's signal stability under surface rotation from -15° to 45°, with quantitative analysis (in Fig. 7(b)) identifying rotation angles where $v_{p-p} < 0.5$. A representative 19° rotation (64° incidence) enables effective concealment while maintaining interception capabilities. The eavesdropping validation at 19° rotation (in Fig. 7(c)) confirms multiple directions with $c_{s-p}<0.5$, indicating persistent security vulnerabilities. Backscattering analysis in Fig. 7(d) shows $S_{180-p}$ values consistently below 0.5 around the optimal rotation angle, demonstrating resistance to transmitter-based anomaly detection.

Now, Based on above analysis, we can say that wallpaper provides maximum signal enhancement at specular direction but creates the highest eavesdropping risk through preserved scattering patterns, curtain offers optimal concealment with minimal performance impact due to low dielectric interactions, while plaster exhibits the most complex behavior with broader energy dispersion from high dielectric constant that marginally improves security but maintains sufficient signal quality for interception.

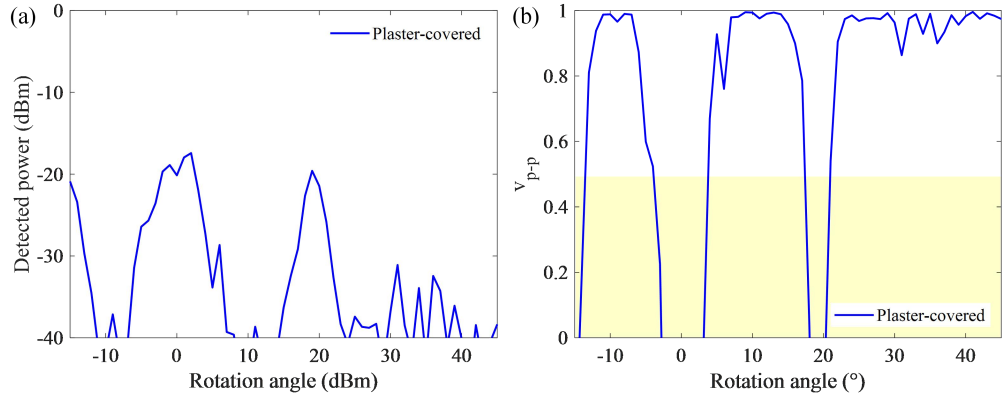

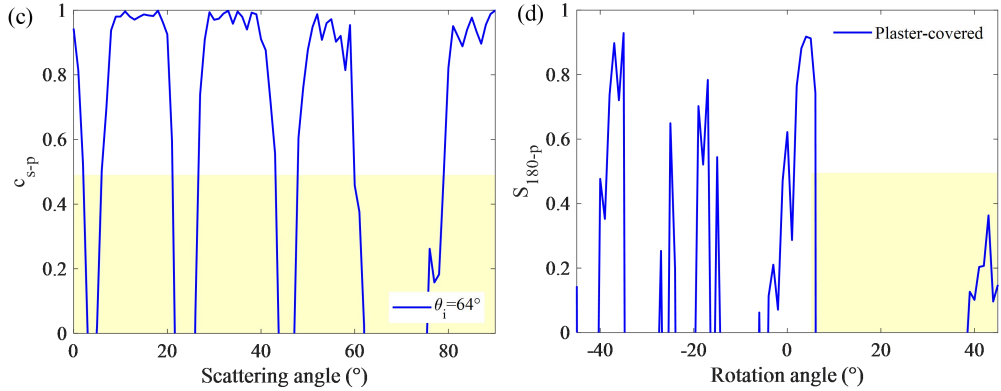

Fig. 7. Plaster security validation at 140 GHz. (a) Bob's signal distribution under different surface rotation angles. (b) Variation parameter $v_{p\text{-}p}$. (c) Secrecy capacity $c_{s\text{-}p}$ at 19° rotation. (d) Backscattering parameter $S_{180\text{-}p}$ under different surface rotation angles with the wall-plaster-covered surface. Yellow regions: $c_{s\text{-}p} < 0.5$.

## 4. BER performance

To further quantify the impact of different coverings on the performance of THz channels, it is necessary to evaluate their bit error rate (BER) performance under the presence of various coverings. A prediction methodology was developed based on measured relationships between scattering angle and BER/SNR characteristics under bare surface conditions [5], where the reference system employed 16-QAM modulation at 176 GHz, closing to our experimental frequency range (113-170 GHz). Validation against published measurements confirms the prediction accuracy (see Fig, S4 in the supplemental document part) of our theretical model under under additive white Gaussian noise (AWGN) conditions, as

$$BER_{16QAM}(\gamma) = \frac{4}{k}\left(1 - \frac{1}{\sqrt{M}}\right)Q\left(\sqrt{\frac{3k}{M-1} \cdot \gamma}\right) \tag{4}$$

where, $M=16$ represents the modulation order as $k = \log_2 M$, $\gamma$ denotes the SNR at each scattering angle, and $Q(\cdot)$ is the Gaussian Q-function. For SNR extraction from measured BER values, the inverse transformation is

$$\gamma = \frac{(Q^{-1}(BER/factor))^2 \cdot (M-1)}{3k} \tag{5}$$

where, $factor = (k/4)/(1-\sqrt{M})$. and where $Q^{-1}(\cdot)$ represents the inverse Q-function. Fig. 8 presents BER distributions across scattering angles for all covering materials at 113 GHz, 140 GHz, and 170 GHz. Angular regions with favorable BER correlate with $c_{s\text{-}p} < 0.5$ conditions (Section 3, supplemental document), confirming eavesdropping feasibility across the frequency band.

The maintained correlation between security vulnerabilities and channel performance demonstrates that covering materials redistribute rather than eliminate eavesdropping threats while preserving sufficient signal quality for information extraction. However, at certain scattering angles, BER cannot reach to the forward error correction (FEC) threshold of $2\times10^{-3}$ due to destructive interference between multiple scattering mechanisms. Specifically, the covering materials create additional electromagnetic interfaces that generate secondary scatterings and transmission losses, which interfere with the primary Bragg scattering from the underlying wavy substrate, resulting in phase mismatches and signal degradation that manifest as elevated BER at specific angular positions where constructive interference conditions are not met.

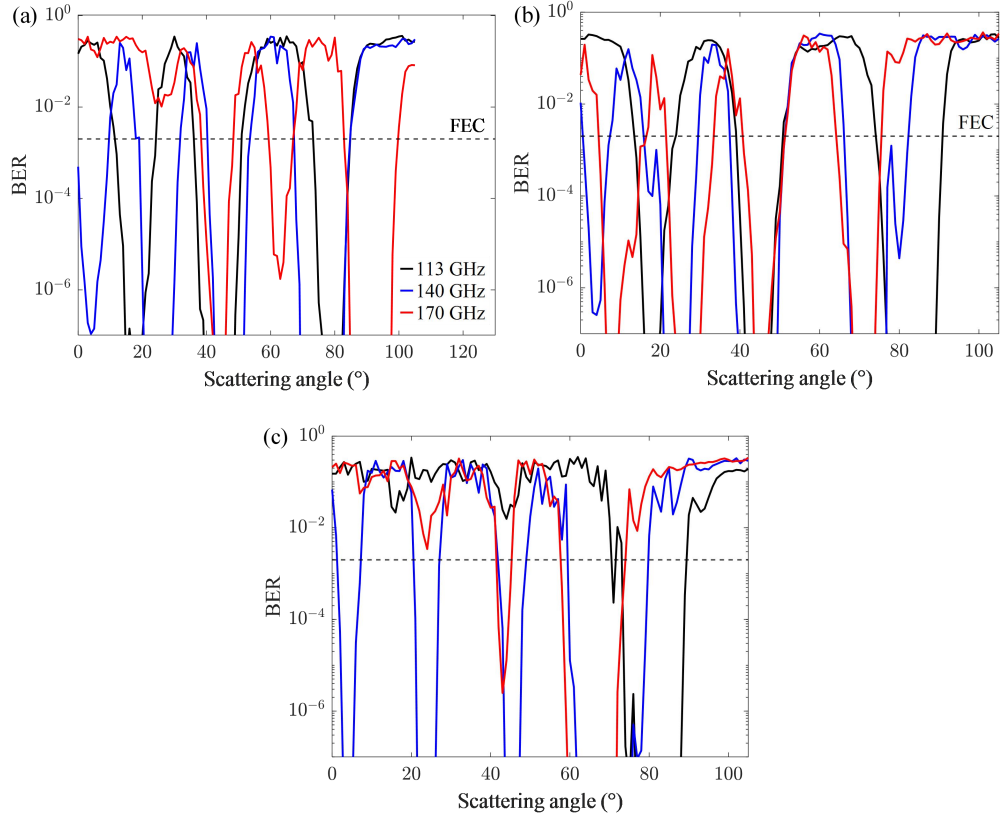

Fig. 8. BER vs. scattering angle for (a) wallpaper, (b) curtain, and (c) wall plaster at 113, 140, and 170 GHz. (b) and (c) keep the same legend with (a).

## 5. Conclusion

This work addresses a critical knowledge gap regarding THz channel behavior and security implications when engineered surfaces are concealed beneath common indoor covering materials - a scenario increasingly relevant for practical deployment of next-generation wireless networks. Through systematic experimental investigation across 113-170 GHz, we characterized the electromagnetic and security properties of MWS covered with three representative indoor materials - wallpaper, curtain, and wall plaster. Our experimental methodology employed dual rotational measurement protocols to comprehensively assess both angular power distribution and incident angle dependencies under NLoS conditions. The results demonstrate that covering materials fundamentally preserve the underlying Bragg scattering characteristics while introducing distinct material-dependent modifications. Wallpaper coverings enhance specular reflection by approximately 10 dB through constructive interference between surface and substrate reflections. Curtain materials exhibit minimal electromagnetic perturbation due to low dielectric properties and porous mesh structure, providing optimal concealment with minimal performance impact. Wall plaster creates smooth interfaces through complete surface filling, resulting in broader energy dispersion and more complex interference patterns.

Critical security analysis reveals that none of these covering approaches eliminate eavesdropping vulnerabilities inherent to THz NLoS systems. Instead, they redistribute threat vectors across different angular domains while maintaining sufficient signal quality for information extraction by potential adversaries. All covering configurations preserve multiple angular positions where normalized secrecy capacity values indicate feasible eavesdropping scenarios, with BER below FEC thresholds at vulnerable angles. Significantly, conventional backscattering detection mechanisms fail to identify surface manipulation attacks, as all tested configurations maintain detection parameters below critical thresholds during optimal eavesdropping conditions. These findings expose fundamental limitations in current physical-layer protection strategies for concealed THz communication systems and underscore the urgent need for enhanced security mechanisms.

**Acknowledgment.** National Natural Science Foundation of China under Grant (62471033); Talent Support Program of Beijing Institute of Technology "Special Young Scholars" (3050011182153).

**Disclosures.** The authors declare no conflicts of interest.

**Data availability.** Data underlying the results presented in this paper are not publicly available at this time but may be obtained from the authors upon reasonable request.